\documentclass[a4paper,11pt]{article}
\pdfoutput=1
\usepackage{jcappub}
\usepackage{mathtools}
\usepackage[T1]{fontenc}
\usepackage[normalem]{ulem}

\graphicspath{{plots/}}

\newcommand{\gs}{g_\star}
\newcommand{\gss}{g_{\star s}}
\newcommand{\Trh}{T_\text{rh}}
\newcommand{\ls}{\lambda_S}
\newcommand{\lhs}{\lambda_{\Phi S}}
\newcommand{\svW}{\langle\sigma v\rangle}
\newcommand{\sv}{\langle\sigma v^3\rangle_{4\to 2}}
\newcommand{\xfo}{x_\text{fo}}
\newcommand{\xrh}{x_\text{rh}}
\newcommand{\arh}{a_\text{rh}}
\newcommand{\afo}{a_\text{fo}}
\newcommand{\Tfo}{T_\text{fo}}

\title{Strongly interacting singlet scalar dark matter during reheating}
\author[a]{Geneviève Bélanger,}
\author[b]{Nicolás Bernal,}
\author[c]{Alexander Pukhov}

\affiliation[a]{LAPTh, CNRS, Université Savoie Mont-Blanc\\
9 Chemin de Bellevue, 74940 Annecy, France}
\affiliation[b]{New York University Abu Dhabi\\
PO Box 129188, Saadiyat Island, Abu Dhabi, United Arab Emirates}
\affiliation[d]{Skobeltsyn Institute of Nuclear Physics, Moscow State University\\
Moscow 119992, Russia}

\emailAdd{belanger@lapth.cnrs.fr}
\emailAdd{nicolas.bernal@nyu.edu}
\emailAdd{alexander.pukhov@gmail.com}

\abstract{We revisit the singlet scalar dark matter model in the presence of a non-standard cosmological history prior to radiation domination. We focus on the regime in which the relic abundance is set by 4-to-2 self-annihilations while the dark and visible sectors remain in kinetic equilibrium, i.e. the standard strongly interacting massive particle (SIMP) framework. In the conventional radiation-dominated cosmology, this realization is not viable, as it requires sub-MeV masses and large quartic couplings in tension with bounds on dark matter self-interactions. We show that this conclusion is significantly modified if freeze-out occurs during non-standard cosmological eras. The altered Hubble expansion rate and the possible non-conservation of the standard model entropy change the freeze-out dynamics, allowing the observed relic density to be achieved with perturbative couplings and consistent with astrophysical constraints. We determine the region where SIMP production dominates over the WIMP mechanism and confront the viable parameter space with current and future direct detection and collider bounds.}
\begin{document}
\begin{flushright}
\end{flushright}
\maketitle

%%%%%%%%%%%%%%%%%%%%%%%%%%%%%%%%%%%%%
\section{Introduction}
%%%%%%%%%%%%%%%%%%%%%%%%%%%%%%%%%%%%%
The existence of dark matter (DM) is firmly established by a wide range of astrophysical and cosmological observations, including galactic rotation curves, gravitational lensing, the cosmic microwave background, and the formation of large-scale structures~\cite{Planck:2018vyg}. Nevertheless, the particle nature of DM and the mechanism responsible for its cosmic abundance remain unknown. Determining how DM was produced in the early Universe is therefore one of the central challenges at the interface of particle physics and cosmology.

A compelling possibility is that DM is a thermal relic. In the standard cosmological scenario, where the Universe is radiation dominated well before Big Bang nucleosynthesis (BBN), the observed relic abundance can be explained if DM was initially in thermal and chemical equilibrium with the standard model (SM) plasma and subsequently froze out. The canonical example is the weakly interacting massive particle (WIMP), whose relic density is set by 2-to-2 annihilations into SM particles~\cite{Arcadi:2024ukq}. Alternatively, if the interaction between the dark and visible sectors is extremely suppressed, DM may never reach equilibrium and can instead be produced via freeze-in (FIMP)~\cite{McDonald:2001vt, Choi:2005vq, Kusenko:2006rh, Petraki:2007gq, Hall:2009bx, Bernal:2017kxu}.

Another well-motivated possibility is that the DM relic density is determined predominantly by number-changing processes within a secluded dark sector. In this context, strongly interacting massive particles (SIMPs)~\cite{Carlson:1992fn, Hochberg:2014dra} arise when the dominant reactions are 3-to-2~\cite{Choi:2015bya, Bernal:2015bla, Bernal:2015lbl, Ko:2014nha, Choi:2017mkk, Chu:2017msm, Bernal:2015ova, Yamanaka:2014pva, Hochberg:2014kqa, Lee:2015gsa, Hansen:2015yaa} or 4-to-2~\cite{Bernal:2015xba, Heikinheimo:2016yds, Bernal:2017mqb, Heikinheimo:2017ofk, Bernal:2018hjm} annihilations between DM particles. These mechanisms typically require sizable DM self-interactions, while keeping non-vanishing portals between the visible and dark sectors, that lead to distinctive phenomenology. However, in minimal realizations and within the standard radiation-dominated cosmological background, the SIMP mechanism is often severely constrained. In particular, in simple Higgs-portal models, achieving the correct relic abundance through self-annihilations while maintaining kinetic equilibrium with the SM bath typically requires light DM masses in the keV to MeV ballpark and large self-couplings, in tension with bounds from astrophysical observations of DM self-interactions~\cite{Hochberg:2014dra, Bernal:2015xba}.

Despite the success of the standard cosmological model, our knowledge of the thermal history of the Universe prior to BBN remains limited. The pre-BBN era could have featured a non-standard expansion history, such as an early matter-dominated phase or a more general background characterized by an equation-of-state parameter $w \neq 1/3$. Such deviations modify the Hubble expansion rate and the evolution of the SM entropy, significantly affecting the dynamics of DM freeze-out~\cite{Allahverdi:2020bys, Batell:2024dsi}. As a result, production mechanisms that are excluded in the standard cosmology may become viable in alternative backgrounds~\cite{Bhatia:2020itt, Bernal:2022wck, Bernal:2023ura, Bernal:2024yhu, Chowdhuryand:2024uvi}.

In this work, we revisit the singlet scalar DM model~\cite{McDonald:1993ex, Burgess:2000yq} --- one of the simplest and most studied extensions of the SM --- in the presence of a non-standard cosmological history prior to radiation domination. We focus on the regime in which the DM relic abundance is determined by 4-to-2 self-annihilations while the dark sector remains in kinetic equilibrium with the SM bath, i.e., the standard SIMP setup. We show that, contrary to the conventional radiation-dominated case, the SIMP mechanism becomes fully viable if chemical freeze-out occurs during an early matter-dominated era or, more generally, in a non-standard cosmological background.

We derive analytical expressions for the freeze-out temperature and relic abundance in both standard and non-standard cosmologies, and validate our results numerically with a modified version of \texttt{micrOMEGAs}~\cite{Alguero:2023zol, Belanger:2026asz}. We demonstrate that, during non-standard cosmological eras, the observed DM abundance can be achieved with perturbative quartic couplings and masses ranging from sub-GeV to ultra-heavy scales. Importantly, the resulting parameter space is compatible with bounds from DM self-interactions and BBN. We further determine the minimal Higgs-portal coupling required to maintain kinetic equilibrium at freeze-out, and identify the region where the SIMP mechanism dominates over standard WIMP production. Finally, we confront the viable parameter space with current limits from direct detection experiments and collider measurements, and discuss the prospects for future facilities such as DARWIN/XLZD, the HL-LHC, and FCC-ee.

The paper is organized as follows. In Section~\ref{sec:framework}, we introduce a general parameterization of non-standard cosmological backgrounds and present the singlet scalar DM model. In Section~\ref{sec:SIMP}, we analyze the dynamics of 4-to-2 SIMP freeze-out during reheating and derive the corresponding relic abundance. We also study perturbativity, self-interaction constraints, the conditions for kinetic equilibrium, and discuss experimental constraints and future prospects. We conclude in Section~\ref{sec:concl}.

%%%%%%%%%%%%%%%%%%%%%%%%%%%%%%%%%%%%%
\section{The Set-up} \label{sec:framework}
%%%%%%%%%%%%%%%%%%%%%%%%%%%%%%%%%%%%%
In this section, we first present a general parameterization of the cosmological background. We then introduce one of the simplest particle physics models of DM: the singlet scalar DM model.

%%%%%%%%%%%%%%%%%%%%%%%%%%%%%%%%%%%%%
\subsection{Beyond standard cosmology}
%%%%%%%%%%%%%%%%%%%%%%%%%%%%%%%%%%%%%
Here we consider the possibility of a non-standard cosmological history prior to Big Bang Nucleosynthesis (BBN)~\cite{Allahverdi:2020bys, Batell:2024dsi}, in order to account for potential uncertainties in the evolution of the early Universe. In particular, deviations from the standard cosmological scenario can significantly modify the naively expected SIMP parameter space, as we show below.

Let us assume that during cosmic reheating after the end of cosmic inflation (or more generally before the onset of the SM radiation-dominated era) the Universe is characterized by an equation-of-state parameter $w$, and that the SM temperature scales as a power law with the scale factor. We denote by $\Trh$ and $\arh$ the SM temperature and the cosmic scale factor at the onset of radiation domination, respectively. The Hubble expansion rate $H$ as a function of the scale factor $a$ can then be written as~\cite{Bernal:2024yhu, Bernal:2024jim, Bernal:2024ndy, Bernal:2025fdr, Banik:2025olw, Bernal:2025fcl, Belanger:2025ack}
\begin{equation} \label{eq:H}
    H(a) \simeq H(\Trh) \times
    \begin{dcases}
        \left(\frac{\arh}{a}\right)^\frac{3(1+w)}{2} &\text{ for } a \leq \arh\,,\\
        \left(\frac{\arh}{a}\right)^2 &\text{ for } \arh \leq a\,,
    \end{dcases}
\end{equation}
while the SM bath temperature evolves as
\begin{equation} \label{eq:T}
    T(a) \simeq \Trh \times
    \begin{dcases}
        \left(\frac{\arh}{a}\right)^\alpha &\text{ for } a \leq \arh\,,\\
        \left(\frac{\arh}{a}\right) &\text{ for } \arh \leq a\,.
    \end{dcases}
\end{equation}
Typically $\alpha > 0$. However, in certain scenarios, it is possible to have $\alpha \leq 0$, in which case the temperature does not decrease before $\arh$~\cite{Co:2020xaf}. During the SM radiation-dominated era in general and at $T = \Trh$ in particular, the Hubble expansion rate is given by
\begin{equation}
    H(T) = \frac{\pi}{3} \sqrt{\frac{\gs(T)}{10}}\, \frac{T^2}{M_P}\,,
\end{equation}
where $\gs(T)$ corresponds to the number of relativistic degrees of freedom that contribute to the SM energy density and $M_P \simeq 2.4 \times 10^{18}$~GeV is the reduced Planck mass.

Here, we comment on the parametrization of reheating in Eqs.~\eqref{eq:H} and~\eqref{eq:T}. The most commonly studied scenario involves a heavy inflaton whose energy density redshifts as non-relativistic matter, corresponding to an equation-of-state parameter $\omega = 0$. In this case, the inflaton decays perturbatively into a pair of SM particles, leading to a temperature scaling exponent $\alpha = 3/8$~\cite{Giudice:2000ex}. This scenario can be realized in viable inflationary models, including Starobinsky inflation~\cite{Starobinsky:1980te} and polynomial inflation~\cite{Drees:2021wgd, Bernal:2021qrl, Drees:2022aea, Bernal:2024ykj}. However, other reheating histories are also possible. For instance, in $\alpha$-attractor inflation models~\cite{Kallosh:2013hoa, Kallosh:2013maa}, the inflaton may oscillate around the minimum of a monomial potential, $V(\phi) \propto \phi^n$ with $n \geq 2$, during reheating. The corresponding equation-of-state parameter is then $\omega = (n-2)/(n+2)$~\cite{Turner:1983he}. The value of $\alpha$ also depends on how the inflaton transfers its energy to radiation, and in particular on the structure of the inflaton-matter couplings~\cite{Co:2020xaf, Garcia:2020wiy, Xu:2023lxw, Barman:2024mqo}. For inflaton decays into bosons, one finds $\alpha = 3/[2(n+2)]$, whereas decays into fermions give $\alpha = 3(n-1)/[2(n+2)]$ for $n<7$ and $\alpha = 1$ for $n>7$. If reheating proceeds through inflaton annihilation into bosons via contact interactions, then $\alpha = 9/[2(n+2)]$ for $n \geq 3$. In scenarios where reheating occurs through $s$-channel annihilation mediated by a light scalar, resonant effects can instead lead to $\alpha = 3(7-2n)/[2(n+2)]$ for bosonic final states and $\alpha = 3(5-n)/[2(n+2)]$ for fermionic final states~\cite{Barman:2024mqo}. If the mediator is massive, inflaton annihilation into a pair of fermions gives $\alpha = 1$~\cite{Barman:2024mqo}. Reheating can also occur at constant temperature, corresponding to $\alpha = 0$~\cite{Co:2020xaf, Barman:2022tzk, Chowdhury:2023jft, Cosme:2024ndc}. If the inflaton energy density is diluted faster than free radiation, namely for $\omega > 1/3$, the inflaton need not decay or annihilate completely; in this case one may have $\alpha = 1$, as in kination scenarios~\cite{Spokoiny:1993kt, Ferreira:1997hj}. Finally, values $\alpha > 1$ are also possible, with $\alpha = 2$ arising, for example, for non-relativistic particles in kinetic equilibrium. In the following, we will focus on the case of early-matter domination ($\omega = 0$ with $\alpha = 3/8$), and two scenarios of radiation domination ($\omega = 1/3$): $\alpha = 1/4$ and $\alpha = 3/4$.

%%%%%%%%%%%%%%%%%%%%%%%%%%%%%%%%%%%%%
\subsection{Singlet scalar DM}
%%%%%%%%%%%%%%%%%%%%%%%%%%%%%%%%%%%%%
The singlet scalar model~\cite{McDonald:1993ex, Burgess:2000yq} is one of the minimal extensions of the SM that provides a viable candidate for DM. In addition to the particle content of the SM, the model introduces a real scalar field $S$. This field is a singlet under the SM gauge group and is assumed to be odd under a $\mathbb{Z}_2$ symmetry, which guaranties its stability.

The most general renormalizable scalar potential consistent with these symmetries is given by
\begin{equation}
    V = \mu_\Phi^2\, |\Phi|^2 + \lambda_\Phi\, |\Phi|^4 + \mu_S^2\, S^2 + \ls\, S^4 + \lhs\, |\Phi|^2\, S^2\,,
\end{equation}
where $\Phi$ is the SM-like Higgs doublet. We require that electroweak symmetry breaking proceeds as in the SM, such that the Higgs acquires a non-vanishing vacuum expectation value $v_\Phi = 246$~GeV, while the singlet does not, $\langle S\rangle = 0$. The latter condition preserves the $\mathbb{Z}_2$ symmetry after symmetry breaking and ensures the stability of the DM candidate. The mass of the singlet is given by $m^2 = \mu_S^2 + \frac12\, \lhs^2\, v_\Phi^2$.

\begin{figure}[t!]
    \def\sepf{0.496}
    \centering
    \includegraphics[width=\sepf\columnwidth]{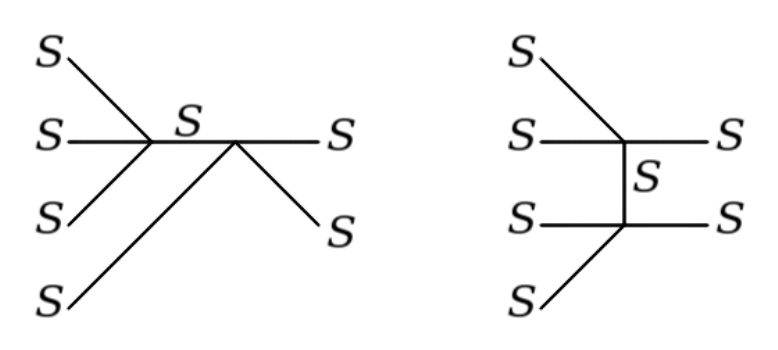}
    \caption{Leading processes for the 4-to-2 DM annihilation, in the limit of small $\lhs$.}
    \label{fig:4-2}
\end{figure}

In general, the evolution of the DM number density $n$ is governed by the Boltzmann equation
\begin{equation} \label{Boltzmann:eqn}
    \frac{dn}{dt} + 3\, H\, n = -\svW \left[n^2 - n_\text{eq}^2\right] -\sv \left[n^4 - n^2\, n_\text{eq}^2\right],
\end{equation}
where $H$ is the Hubble rate and $n_\text{eq}$ denotes the equilibrium number density of DM particles. The first term on the right-hand side corresponds to DM annihilation into SM particles (and the inverse process), with $\svW$ the thermally averaged 2-to-2 annihilation cross section. The second term accounts for number-changing self-interactions in the dark sector (e.g. 4-to-2 processes), characterized by the thermally averaged effective ``cross section'' $\sv$ (with mass dimension $-8$), as illustrated in Fig.~\ref{fig:4-2}. This is the leading number-changing process within the dark sector: 2-to-2 DM elastic scatterings conserve the DM number density, while 3-to-2 processes are forbidden by the $\mathbb{Z}_2$ symmetry.
%%%%%%%%%%%%%%%%%%%%%%%%%%%%%%%%%%%%%%%%%%%%%%%%%%%
%%%%%%%%%%%%%%%%%%%%%%%%%%%%%%%%%%%%%%%%%% 

In the absence of DM self-interactions (i.e., for $\ls = 0$), large values of the portal coupling, $\lhs \sim \mathcal{O}(10^{-1}$), lead to the standard WIMP solution~\cite{McDonald:1993ex, Burgess:2000yq}, while very suppressed couplings, $\lhs \sim \mathcal{O}(10^{-11})$, give rise to the FIMP solution~\cite{Yaguna:2011qn, Bernal:2018kcw}. Interestingly, in scenarios with low reheating temperatures, there exists a smooth transition between these two apparently disconnected regimes~\cite{Bernal:2018ins, Cosme:2023xpa, Silva-Malpartida:2023yks, Arcadi:2024wwg, Silva-Malpartida:2024emu, Lebedev:2024mbj}.

In the standard cosmological scenario, the SIMP solution—where the visible and dark sectors share the same temperature—is not viable in this model. Achieving the correct relic abundance would require a quartic coupling $\ls \sim \mathcal{O}(1)$ and a DM mass $m \sim \mathcal{O}(100)$~keV~\cite{Bernal:2015xba}, which is in tension with bounds on DM self-interactions from observations of the Bullet Cluster~\cite{Clowe:2003tk, Randall:2008ppe, Markevitch:2003at, Bernal:2015xba}. However, DM production through 4-to-2 annihilations becomes viable if the DM mass is in the MeV–TeV range and the dark sector is much colder than the SM bath. This temperature hierarchy can be dynamically generated through a very small portal coupling $\lhs \sim \mathcal{O}(10^{-12})$, in a freeze-in–like setup~\cite{Bernal:2015xba}.

In the next section, we present a novel solution within the singlet scalar DM framework. We demonstrate that the standard SIMP mechanism—namely, DM production via 4-to-2 annihilations while the dark sector remains in kinetic equilibrium with the SM bath—can become viable in the presence of a non-standard cosmological background.

%%%%%%%%%%%%%%%%%%%%%%%%%%%%%%%%%%%%%
\section{Strongly interacting singlet scalar DM during reheating} \label{sec:SIMP}
%%%%%%%%%%%%%%%%%%%%%%%%%%%%%%%%%%%%%
Let us begin by assuming that the DM is in kinetic equilibrium with the SM bath, so that both sectors share a common temperature $T$. This condition is satisfied, provided that the portal coupling  $\lhs$ is not too suppressed. In addition, we focus on the regime in which the DM relic abundance is determined by dark-sector self-interactions. This can occur when the portal coupling is sufficiently small to prevent standard 2-to-2 annihilations into SM states from dominating the freeze-out process. The region of parameter space where these two conditions are simultaneously satisfied will be quantified in the following. For simplicity, we temporarily set $\lhs = 0$ and concentrate on the dynamics induced solely by DM self-interactions.

\begin{figure}[t!]
    \def\sepf{0.496}
    \centering
    \includegraphics[width=\sepf\columnwidth]{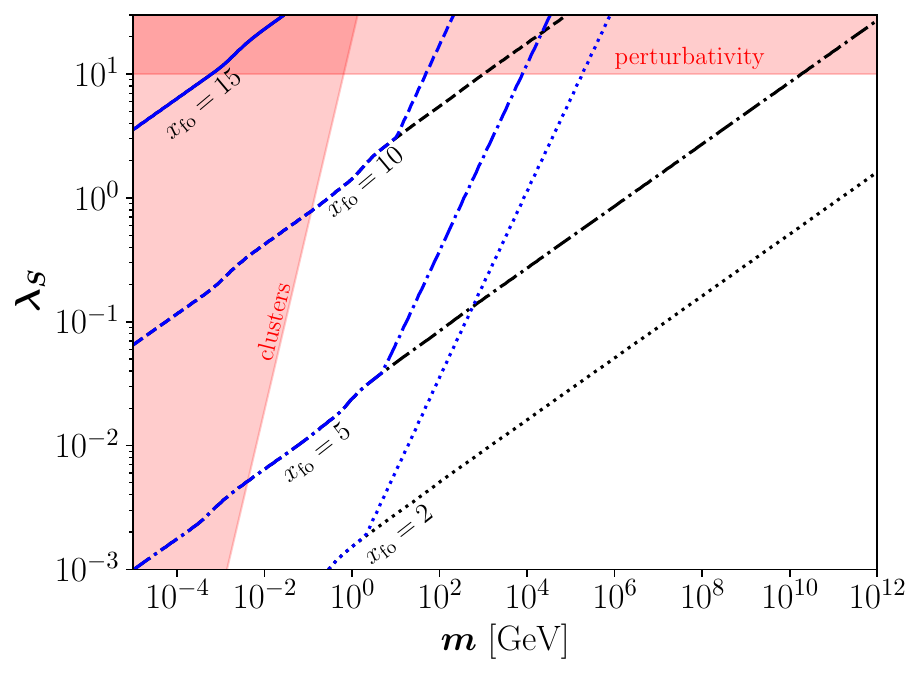}
    \caption{Contours for the freeze out temperature $\xfo \equiv m/\Tfo$ in the plane [$m, \ls$], for $\xfo = 2$, 5, 10, and 15. The black lines correspond to the standard cosmological scenario, while the blue lines correspond to an early matter domination ($w = 0$ with $\alpha = 3/8$) with $\Trh = 1$~GeV.}
    \label{fig:xfo}
\end{figure}

In this scenario, DM is also in {\it chemical} equilibrium with itself through 4-to-2 reactions. The DM freeze out temperature $\Tfo$ can be implicitly estimated by the equality between the Hubble expansion and the interaction rates $H(\Tfo) = \sv \times n_\text{eq}^3(\Tfo)$, where, in the non-relativistic limit, the DM number density at equilibrium can be approximated using the Maxwell-Boltzmann statistics by
\begin{equation}
    n_\text{eq}(T) \simeq \left(\frac{m\, T}{2 \pi}\right)^{3/2} e^{-\frac{m}{T}},
\end{equation}
and, again in the non-relativistic limit, the 4-to-2  DM annihilation cross section is given by
\begin{equation} \label{eq:sv}
    \sv \simeq \frac{54\sqrt{3}}{\pi}\, \frac{\ls^4}{m^8}\,.
\end{equation}
A detailed derivation of $\sv$ can be found in Appendix~\ref{sec:app}. It follows that
\begin{equation}
    \xfo \simeq
    \begin{dcases}
        \frac56\, \mathcal{W}_0\left[\frac{9\, \ls^{8/5}}{2 \pi^{13/5}\, \gs^{1/5}} \left(\frac{3}{10}\right)^{4/5} \left(\frac{M_P}{m}\right)^{2/5}\right] &\text{for } \xrh \leq \xfo\,,\\
        \frac{3 \alpha - w - 1}{2\,\alpha}\, \mathcal{W}_k\left[\frac{2\,\alpha}{3 \alpha - w - 1}\, \xrh \left(\frac{5\times 27^3\, \ls^8}{2^{14} \pi^{13} \gs}\, \frac{M_P^2}{m^2\, \xrh^5}\right)^\frac{\alpha}{3(3 \alpha - w - 1)}\right] &\text{for } \xfo \leq \xrh\,,
    \end{dcases}
\end{equation}
where $\mathcal{W}_k(x)$ is the Lambert $W$ function, with $k = 0$ if $\alpha/(3 \alpha - w -1) > 0$ or $k = -1$ if $\alpha/(3 \alpha - w -1) < 0$. As expected, the two cases coincide when $w = 1/3$ with $\alpha = 1$. Figure~\ref{fig:xfo} shows the contours for $\xfo = 2$, 5, 10, and 15: the black lines correspond to the standard cosmological scenario, while the blue lines correspond to an early matter domination ($w = 0$ and $\alpha = 3/8$) with $\Trh = 1$~GeV. It can be seen that large regions of the parameter space, corresponding to very high DM masses, are compatible with a non-relativistic freeze-out ($\xfo > 2$). In addition, for non-standard cosmological scenarios, the extra component in the Hubble expansion rate induces an early decoupling and therefore a decrease in $\xfo$.
%%%%%%%%%%%%%%%%%%%%%%%%%%%%%%%%%%%%%%%%%%%%%%%%%%%
%%%%%%%%%%%%%%%%%%%%%%%%%%%%%%%%%%%%%%%%%% 

In Fig.~\ref{fig:xfo}, we also overlay the perturbativity limit, $\ls \lesssim 10$, as well as the constraints from DM self-interactions. Indeed, a sizable quartic coupling $\ls$ can induce very efficient 2-to-2 DM elastic scatterings, potentially in conflict with astrophysical observations. In the limit of a small Higgs portal coupling and at low velocities, the ratio of the DM self-interaction cross section to the DM mass is given by~\cite{Bernal:2015xba}
\begin{equation}
    \frac{\sigma_{SS}}{m} \simeq \frac{9}{8 \pi}\, \frac{\ls^2}{m^3}\,.
\end{equation}
Too strong self-interactions could lead to gravothermal collapse in the cores of DM halos~\cite{Vogelsberger:2012ku}. In particular, observations of the Bullet Cluster constrain the DM self-interaction cross section to be below 1.25~cm$^2$/g at 68\% CL~\cite{Markevitch:2003at, Clowe:2003tk, Randall:2008ppe}. More recently, analysis of galaxy cluster collisions has yielded a stronger bound $\sigma_{SS}/m < 0.47$~cm$^2$/g at 95\% CL~\cite{Harvey:2015hha}, and is represented in Fig.~\ref{fig:xfo} by the diagonal red band.

We now turn to the evolution of the DM number density. For {\it small} values of the portal coupling $\lhs$, the first term (proportional to $\lhs^2$) on the right-hand side of Eq.~\eqref{Boltzmann:eqn} is subdominant with respect to the second term (independent of $\lhs$). Therefore, Eq.~\eqref{Boltzmann:eqn} can be conveniently rewritten in terms of the comoving number density $N \equiv n\, a^3$ as
\begin{equation} \label{Boltzmann:eqn2}
    \frac{dN}{da} \simeq -\frac{\sv}{a^{10}\, H} \left[N^4 - N^2\, N_\text{eq}^2\right],
\end{equation}
where $N_\text{eq} \equiv n_\text{eq}\, a^3$. Anticipating the discussion, we note that the strength of the portal coupling $\lhs$ is restrained by demanding that the SIMP solution dominates over the WIMP solution, and the requirement of kinetic equilibrium for DM.

Assuming that DM is chemically in equilibrium with itself until freeze-out at $a = \afo$, and that at least until that point there is kinetic equilibrium between DM and the SM, Eq.~\eqref{Boltzmann:eqn2} can be solved analytically, integrating from $\afo$, and realizing that after freeze-out $N_\text{eq} \ll N$. It follows that
\begin{equation}
    N(a)^3 \simeq \frac{5 - w}{2}\, \frac{H(\afo)}{\sv}\, \afo^9 \left[1 - \left(\frac{\afo}{a}\right)^\frac{3(5-w)}{2}\right]^{-1}.
\end{equation}
This implies that if DM freeze-out occurs during the standard cosmological scenario ($w = 1/3$ and $\alpha = 1$), at late times, when $a = a_0 \gg \afo$,
\begin{equation} \label{eq:SC}
    Y_0 \equiv \frac{N(a_0)}{a_0^3\, s(a_0)}\simeq \frac{15}{2\, \gss(\Tfo)} \left[\frac{21}{\pi^5} \sqrt{\frac{\gs(\Tfo)}{10}}\, \frac{1}{M_P\, \sv} \left(\frac{\xfo}{m}\right)^7\right]^{1/3},
\end{equation}
where $\xfo \equiv m/\Tfo$, and
\begin{equation}
    s(T) = \frac{2 \pi^2}{45}\, \gss(T)\, T^3
\end{equation}
is the SM entropy density, with $\gss(T)$ the relativistic degrees of freedom contributing to the SM entropy.

In contrast, if freeze-out occurs during reheating, Eq.~\eqref{Boltzmann:eqn2} has to be integrated in two steps: between $\Tfo$ and $\Trh$ (with general values for $w$ and $\alpha$), then from $\Trh$ (taking $w = 1/3$ and $\alpha = 1$). It follows that
\begin{align} \label{eq:NSC}
    Y_0 &\simeq \frac{15}{2\, \gss(\Trh)} \left[\frac{21}{\pi^5} \sqrt{\frac{\gs(\Trh)}{10}}\, \frac{1}{M_P\, \sv} \left(\frac{\xrh}{m}\right)^7\right]^\frac13 \nonumber\\
    &\qquad\qquad\qquad \times \left[1 + \frac{14}{3(5-w)} \left(\left(\frac{\xrh}{\xfo}\right)^\frac{3(5-w)}{2 \alpha} - 1\right)\right]^{-\frac13},
\end{align}
where $\xrh \equiv m/\Trh$. In the second squared bracket, the second term corresponds to DM production during reheating, while the first term corresponds to the contribution after reheating, which can be compared to Eq.~\eqref{eq:SC}. As expected, Eq.~\eqref{eq:NSC} reduces to Eq.~\eqref{eq:SC} when $\xrh = \xfo$ or when $w = 1/3$ with $\alpha = 1$.

\begin{figure}[t!]
    \def\sepf{0.496}
    \centering
    \includegraphics[width=\sepf\columnwidth]{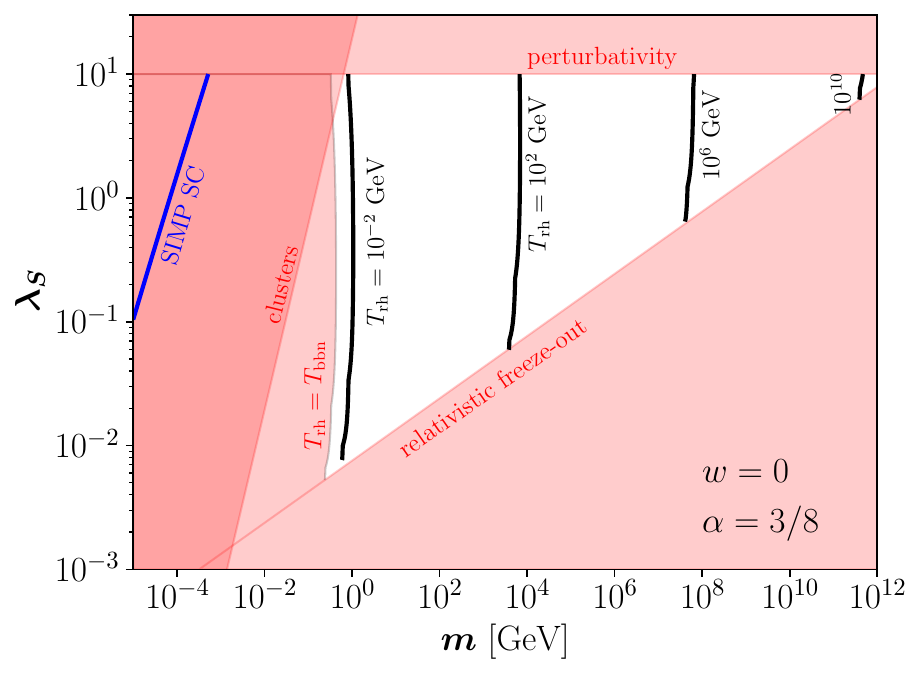}
    \includegraphics[width=\sepf\columnwidth]{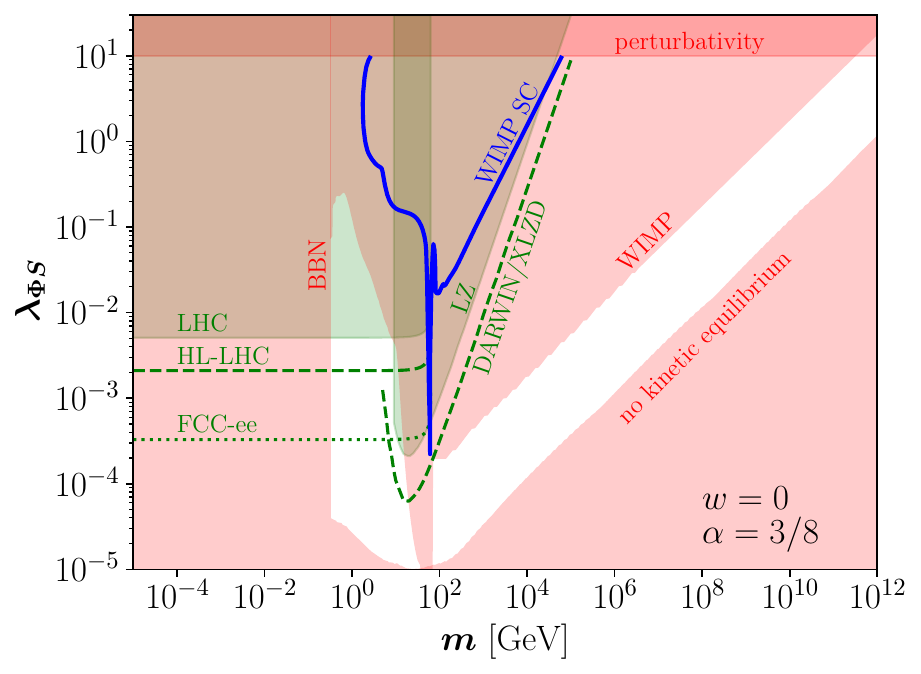}
    \caption{Parameter space that fits the entire DM relic abundance through the SIMP mechanism, for $\Trh = 10^{-2}$~GeV, $10^2$~GeV, $10^6$~GeV and $10^{10}$~GeV, assuming an early matter dominated era ($w=0$ and $\alpha = 3/8$). The blue lines correspond to the standard cosmological scenario. Left panel: The red area are in tension with the Bullet cluster, perturbativity, BBN, or require a relativistic freeze-out. Right panel: The green area are in tension with direct DM searches and the invisible decay of the Higgs while in the red area  DM does not reach kinetic equilibrium with the SM, gives rise to a WIMP solution or is ruled out by BBN.}
    \label{fig:m-ls}
\end{figure}

To match the entire observed DM relic density, it is required that
\begin{equation}
    m\, Y_0 = \frac{\Omega h^2\, \rho_c}{s_0\, h^2} \simeq 4.3 \times 10^{-10}~\text{GeV},
\end{equation}
where $Y_0$ is the asymptotic value of the DM yield at low temperatures, $s_0 \simeq 2.89 \times 10^3$~cm$^{-3}$ is the present entropy density~\cite{ParticleDataGroup:2024cfk}, $\rho_c \simeq 1.05 \times 10^{-5}~h^2$~GeV/cm$^3$ is the critical energy density of the Universe, and $\Omega h^2 \simeq 0.12$ is the observed DM relic abundance~\cite{Planck:2018vyg}. In the left panel of Fig.~\ref{fig:m-ls}, the blue line shows the parameter space that fits the entire DM relic abundance for a freeze-out occurring during the standard cosmological scenario. It could be viable for $m \sim \mathcal{O}$(keV) and $\ls \sim 1$; however, it is deep in a region excluded by the Bullet cluster. It is interesting to note that this is no longer the case when DM is mainly produced during an early matter dominated era ($w = 0$ and $\alpha = 3/8$), as shown with black lines for $\Trh = 10^{-2}$~GeV, $10^2$~GeV, $10^6$~GeV, and $10^{10}$~GeV. For the singlet scalar DM model, the SIMP solution is viable for $10^{-1}$~GeV $\lesssim m \lesssim 10^{12}$~GeV, with perturbative couplings. Figure~\ref{fig:m-ls} also overlays the bounds from the Bullet cluster, perturbativity $\ls \lesssim 10$, relativistic freeze-out $\xfo \geq 2$ and BBN $\Trh \geq T_\text{bbn} \simeq 4$~MeV~\cite{Sarkar:1995dd, Kawasaki:2000en, Hannestad:2004px, Barbieri:2025moq}. This figure was produced with the former analytical expressions and was numerically cross-checked with an implementation of the model in \texttt{micrOMEGAs}.
%%%%%%%%%%%%%%%%%%%%%%%%%%%%%%%%%%%%%%%%%%%%%%%%%%%
%%%%%%%%%%%%%%%%%%%%%%%%%%%%%%%%%%%%%%%%%% 

It is important to emphasize that the SIMP paradigm relies on the fact that {\it kinetic equilibrium} between the dark and visible sectors is maintained during DM freeze-out; that is, that the two sectors share the same temperature. This can be granted if the {\it elastic} scattering rate between DM and SM particles is larger than the Hubble expansion rate, at $T = \Tfo$:
\begin{equation}
    H(\Tfo) < \Gamma_\text{el}(\Tfo) \equiv \sum_{i \in \text{SM}} n_{\text{eq}, i}(\Tfo) \times \langle\sigma_\text{el} v\rangle_i(\Tfo)\,,
\end{equation}
where the sum is performed on all SM particles $i$ with equilibrium number density $n_{\text{eq}, i}(T)$. The thermally-averaged elastic scattering cross section DM-SM $\langle\sigma_\text{el} v\rangle_i$ was calculated numerically, integrating the squared matrix elements extracted from {\tt CalcHEP}~\cite{Belyaev:2012qa}. In the right-hand panel of Fig.~\ref{fig:m-ls}, we show the {\it minimal} portal coupling $\lhs$ required to ensure the kinetic equilibrium between the two sectors at the time of chemical freeze-out. For lower values of $\lhs$ (corresponding to the lower red area denoted by `no kinetic equilibrium') DM is not in kinetic equilibrium at the moment of its freeze-out. For high DM masses $m \gg m_W$, the kinetic equilibrium is dominated by DM scatterings with $W$ bosons. Portal couplings as low as $\lhs \simeq 10^{-5}$ are able to maintain equilibrium if $m \simeq 80$~GeV and $\lhs \propto \sqrt{m}$ for higher masses. However, for masses $m < m_W$, the freeze-out occurs at temperatures at which the abundance of the $W$ boson is Boltzmann suppressed, and therefore kinetic equilibrium has to be maintained through scatterings with light fermions. The suppression by small Yukawa couplings in these processes has to be counterbalanced by an increase in the portal coupling. In this panel, the colored region at low masses $m \lesssim 10^{-1}$~GeV comes from the BBN bound in the left-hand panel.

Having a non-zero portal coupling $\lhs$ implies that DM can also be produced through the WIMP mechanism, as a couple of SM particles can annihilate into a pair of DM particles.\footnote{FIMP production is not viable, as DM is in chemical equilibrium with the SM plasma.} As we focus on SIMP production, we demand that the freeze-out of the 4-to-2 annihilations occurs {\it after} the WIMP freeze-out, so that the WIMP mechanism will have no effect on the final DM abundance.\footnote{As DM is assumed to reach chemical equilibrium, the reaction that fixes its abundance is the last to go out of equilibrium. That is why we require the freeze-out of 2-to-2 interactions (that is, the WIMP freeze out) to occur before the freeze-out of 4-to-2 interactions (that is, the SIMP freeze out).} This implies an {\it upper} bound on $\lhs$, and corresponds to the upper red region `WIMP' on the right-hand panel of Fig.~\ref{fig:m-ls}.\footnote{Even if a large portion of this region, in which DM is produced through the WIMP mechanism during reheating, is viable we disregard it as here we are focusing on the SIMP solution.} The strong dip at $m \simeq m_h/2 \simeq 62$~GeV corresponds to the Higgs resonant DM production. For the WIMP production, the freeze-out temperature in this non-standard cosmological scenario was computed with a modified version of \texttt{micrOMEGAs}.

In the right-hand panel of Fig.~\ref{fig:m-ls} we also show the actual constraints from DM direct detection, in particular LZ~\cite{LZ:2024zvo}, and the projected sensitivity of DARWIN/XLZD~\cite{OHare:2021utq, Baudis:2024jnk}. In addition, for DM lighter than $m_h/2$, the coupling $\lhs$ is strongly constrained from the measurement of the invisible Higgs width at the LHC (Br$_\text{h $\to$ inv} < 10.7\%$)~\cite{ATLAS:2023tkt}. We show the current limit and the prospects for HL-LHC (Br$_\text{h $\to$ inv} < 2\%$)~\cite{Okawa:2013hda} and FCC-ee (Br$_\text{h $\to$ inv} < 0.05\%$)~\cite{mehta_2025_7hbn8-3d233}. For reference, we also overlaid in blue the parameter space required for the WIMP solution to be realized in the standard cosmological scenario (`WIMP SC'), computed with \texttt{micrOMEGAs}.

We emphasize that, in the singlet scalar DM model—and contrary to what happens in the standard cosmological scenario—the SIMP production mechanism becomes fully viable if it takes place during an early matter-dominated era. It can be realized for DM masses in the range $10^{-1}$~GeV $\lesssim m \lesssim 10^{12}$~GeV, quartic coupling $5\times 10^{-3} \lesssim \ls \lesssim 10^1$ and Higgs portal coupling $10^{-5} \lesssim \lhs \lesssim 10^1$ (depending on the mass value), corresponding to the white areas in Fig.~\ref{fig:m-ls}. Parts of the parameter space favored by this scenario are already constrained by direct-detection experiments and precision measurements at the LHC, and will be further probed by future facilities such as DARWIN, the HL-LHC, and FCC-ee.

Finally, we note that the regions labeled ``relativistic freeze-out'', ``WIMP'', and ``no kinetic equilibrium'' in Fig.~\ref{fig:m-ls} are excluded under our assumptions, although they could become viable in modified frameworks. First, relativistic freeze-out is possible provided that DM becomes non-relativistic before matter-radiation equality. Second, the WIMP solution during reheating is largely unconstrained; however, it does not correspond to a SIMP and is therefore not considered here. Third, DM production without kinetic equilibrium can also be viable, but it likewise does not realize the SIMP mechanism. Instead, it corresponds to the Elastically Decoupling DM'' (ELDER) scenario~\cite{Kuflik:2015isi, Kuflik:2017iqs}.

\begin{figure}[t!]
    \def\sepf{0.496}
    \centering
    \includegraphics[width=\sepf\columnwidth]{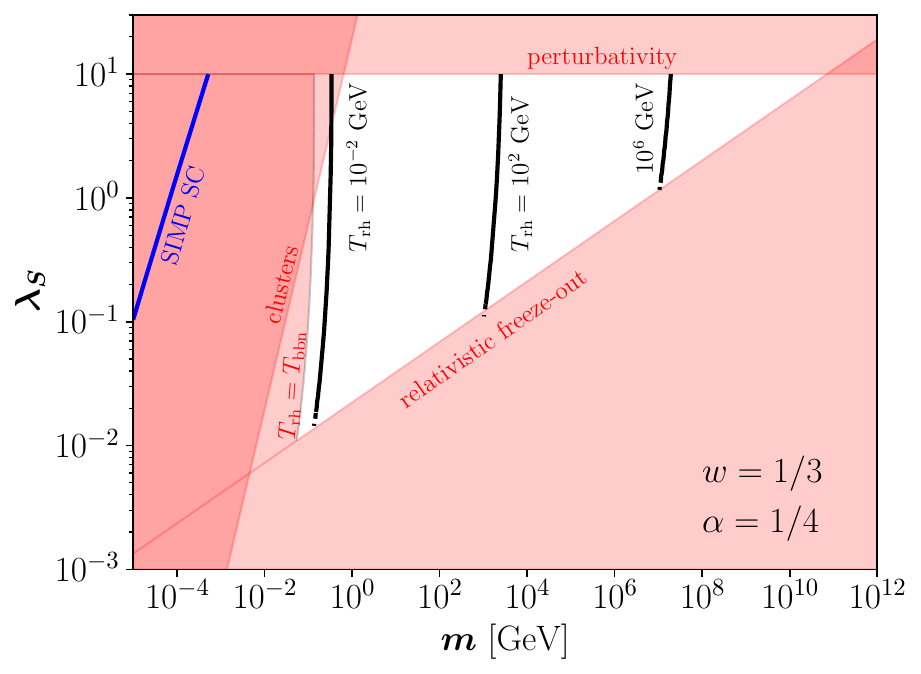}
    \includegraphics[width=\sepf\columnwidth]{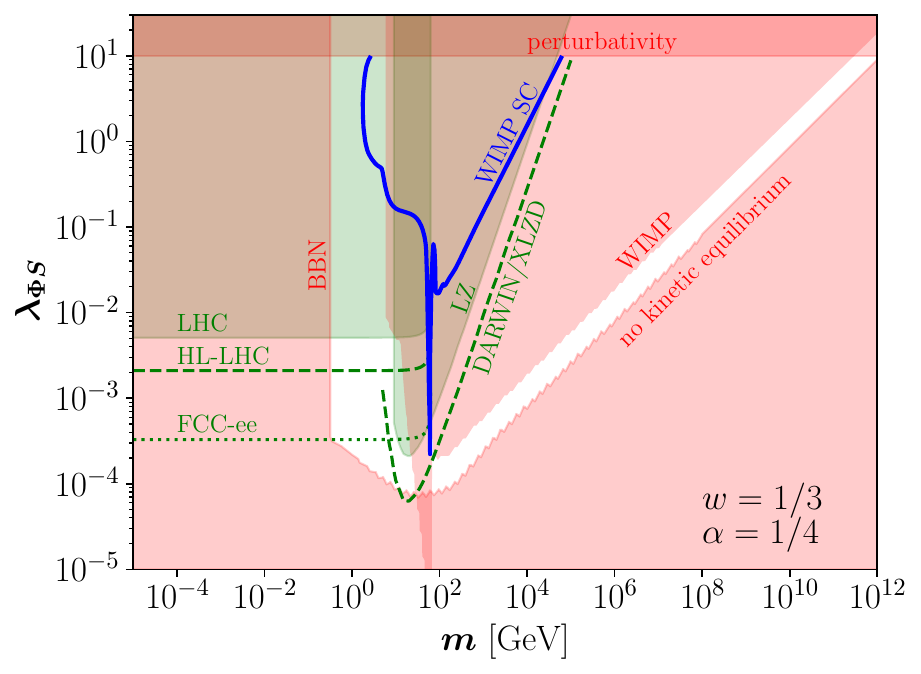}
    \includegraphics[width=\sepf\columnwidth]{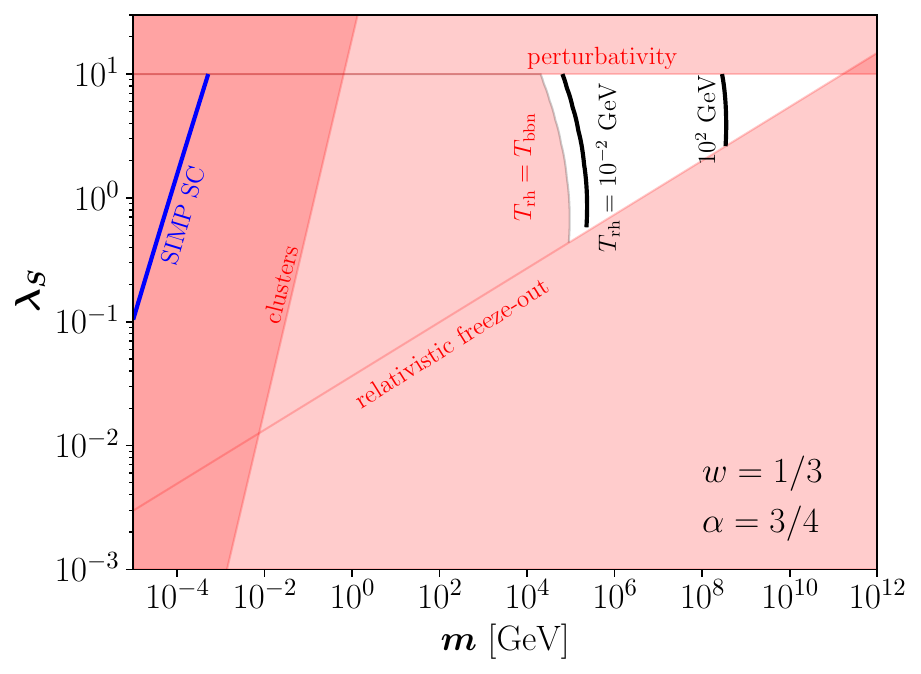}
    \includegraphics[width=\sepf\columnwidth]{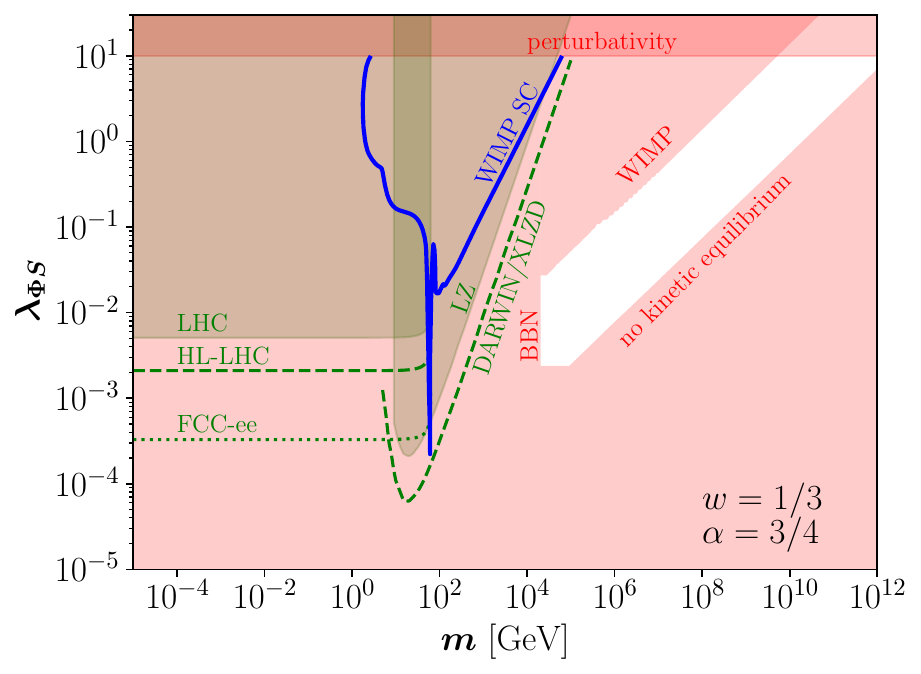}
    \caption{Same as Fig.~\ref{fig:m-ls} but for a radiation-dominated scenario ($w=1/3$) with $\alpha = 1/4$ (top) or $\alpha = 3/4$ (bottom).}
    \label{fig:m-ls-RD}
\end{figure}

The previous conclusion obtained from Fig.~\ref{fig:m-ls} can be generalized to other cosmological scenarios. For completeness, in Fig.~\ref{fig:m-ls-RD} we present similar results but for a radiation-dominated era ($w = 1/3$) with $\alpha = 1/4$ (top) and $\alpha = 3/4$ (bottom). These cosmological scenarios could occur, for example, in the case where the inflaton oscillates in a quartic potential ($w = 1/3$) while perturbatively decaying into a pair of bosons ($\alpha = 1/4$) or fermions ($\alpha = 3/4$). It is worth noticing that the latter case is very constrained. In fact, having $w = 1/3$ and a small injection of entropy because $\alpha$ is close to unity, makes this case close to the standard cosmology and only viable if the DM mass is above the TeV scale.
%%%%%%%%%%%%%%%%%%%%%%%%%%%%%%%%%%%%%%%%%%%%%%%%%%%
%%%%%%%%%%%%%%%%%%%%%%%%%%%%%%%%%%%%%%%%%% 

%%%%%%%%%%%%%%%%%%%%%%%%%%%%%%%%%%%%%%%%%%%%%%%%%%%%%%%%%%
\section{Conclusions} \label{sec:concl}
%%%%%%%%%%%%%%%%%%%%%%%%%%%%%%%%%%%%%%%%%%%%%%%%%%%%%%%%%%
In this work, we have revisited the singlet scalar dark matter (DM) model in the presence of a non-standard cosmological history prior to the onset of radiation domination. We focused on the regime in which the DM relic abundance is determined by 4-to-2 number-changing self-interactions, while the dark and visible sectors remain in kinetic equilibrium during freeze-out --- namely, the standard SIMP framework.

Within conventional radiation-dominated cosmology, this realization of the SIMP mechanism in the singlet scalar model is not viable. Achieving the observed relic abundance typically requires quartic couplings of order unity and sub-MeV DM masses, leading to strong tension with bounds from DM self-interactions, in particular from galaxy cluster observations such as the Bullet Cluster.

We have shown that this conclusion changes dramatically in the presence of a non-standard expansion history. In particular, if chemical freeze-out occurs during an early matter-dominated era, or in more general cosmological backgrounds, the modified Hubble rate and the possible violation of the SM entropy alter the freeze-out dynamics in such a way that the correct relic abundance can be obtained with perturbative quartic couplings and without violating self-interaction constraints. We derived analytical expressions for the freeze-out temperature and relic abundance in both standard and non-standard cosmologies and validated them numerically with a modified version of \texttt{micrOMEGAs}.

In the case of non-standard cosmological eras, the SIMP solution becomes viable over a remarkably broad mass range, typically $10^{-1}~\text{GeV} \lesssim m \lesssim 10^{12}$~GeV, with quartic couplings in the perturbative regime and consistent with astrophysical bounds. We also determined the minimal Higgs-portal coupling required to maintain kinetic equilibrium at freeze-out and identified the upper bound on the portal coupling for which the SIMP mechanism dominates over standard WIMP production.

We confronted the viable parameter space with current experimental constraints, including limits from direct detection experiments and measurements of the invisible Higgs width at the LHC. We found that part of the parameter space is already constrained, while future facilities such as DARWIN, the HL-LHC, and FCC-ee will probe significant additional regions. 

Our results demonstrate that conclusions about the viability of specific DM production mechanisms can depend crucially on assumptions about the pre-BBN cosmological history. In particular, the singlet scalar model --- often considered incompatible with the standard SIMP scenario --- can successfully realize it in the presence of an early matter-dominated era. This highlights the importance of jointly exploring particle physics models and non-standard cosmological histories when assessing the viability of DM scenarios.

%%%%%%%%%%%%%%%%%%%%%%%%%%%%%%%%%%%%%%%%%%%
\acknowledgments
%%%%%%%%%%%%%%%%%%%%%%%%%%%%%%%%%%%%%%%%%%%
NB received funding from the grants PID2023-151418NB-I00 funded by MCIU/AEI/10.13039 /501100011033/ FEDER and PID2022-139841NB-I00 of MICIU/AEI/10.13039/501100011033 and FEDER, UE. NB thanks the Laboratoire d’Annecy-le-Vieux de Physique Théorique (LAPTh) for its hospitality and CNRS-INP for partial support. This study was conducted within the scientific program of the National Center for Physics and Mathematics, section \#5 ``Particle Physics and Cosmology'' stage 2026-2028.

%%%%%%%%%%%%%%%%%%%%%%%%%%%%%%%%%%%%%%%%%%%
\appendix
%%%%%%%%%%%%%%%%%%%%%%%%%%%%%%%%%%%%%%%%%%%
\section{Interaction rates} \label{sec:app}
%%%%%%%%%%%%%%%%%%%%%%%%%%%%%%%%%%%%%%%%%%%
In this appendix, we compute the interaction rate density $\gamma_{4 \to 2}$ for the 4-to-2  DM annihilation, for particles with momentum $\vec p_i$ and energy $E_i$, with $i = 1, \dots, 6$~\cite{Gondolo:1990dk}:
\begin{align}
	\gamma_{4 \to 2}(T) &\equiv \int \frac{d^3\vec p_1}{(2\pi)^32E_1} \frac{d^3\vec p_2}{(2\pi)^32E_2} \frac{d^3\vec p_3}{(2\pi)^32E_3} \frac{d^3\vec p_4}{(2\pi)^32E_4}\, f(T,\vec p_1)\, f(T,\vec p_2)\, f(T,\vec p_3)\, f(T,\vec p_4) \nonumber \\
	&\quad \times \int\frac{d^3\vec p_5}{(2\pi)^32E_5} \frac{d^3\vec p_6}{(2\pi)^32E_6}\, (2\pi)^4\, \delta^{(4)}\left(p_1 + p_2 + p_3 + p_4 - p_5 - p_6\right) |\mathcal{M}|^2\,,
\end{align}
where $f(T,\vec p)$ is the DM phase-space distribution, $|\mathcal{M}|^2$ is the squared matrix element with the corresponding symmetry factors, and where the possible Bose enhancement for the final state has been ignored as $f \ll 1$.

Several approximations can be done in the {\it non-relativistic limit} for the 4 incoming particles. First, $i)$ the squared matrix element (computed with {\tt CalcHEP}) is constant
\begin{equation}
    |\mathcal{M}|^2 = \frac{24^3\, \ls^4}{m^4}
\end{equation}
and can be pulled out of the integrals. Then, $ii)$ the final state integrals become
\begin{align}
    &\int\frac{d^3\vec p_5}{(2\pi)^32E_5} \frac{d^3\vec p_6}{(2\pi)^32E_6}\, (2\pi)^4\, \delta^{(4)}\left(\vec p_1 + \vec p_2 + \vec p_3 + \vec p_4 - \vec p_5 - \vec p_6\right) \nonumber\\
    &\qquad = \int d\Omega\, \frac{\sqrt{\lambda(s, m^2, m^2)}}{8\, (2 \pi)^2\, s} = \frac{\sqrt{3}}{16\, \pi}\,,
\end{align}
where $\lambda(a,b,c) = a^2 + b^2 + c^2 - 2ab - 2ac - 2bc$, $d\Omega$ is the solid angle and $\sqrt{s} = 4\, m$ is the center-of-mass energy. Finally, $iii)$ the initial state integrals
\begin{align}
    \int \frac{d^3\vec p_i}{(2\pi)^32E_i} f(T,p_i) = \frac{n(T)}{2\, m}\,,
\end{align}
for $i = 1, \dots, 4$. With these three ingredients, the interaction rate density becomes
\begin{equation}
	\gamma_{4 \to 2}(T) = n^4(T)\times \sv\,
\end{equation}
with
\begin{equation} \label{eq:sv2}
    \sv = \frac{54\, \sqrt{3}}{\pi}\, \frac{\ls^4}{m^8}\,,
\end{equation}
with corresponds to Eq.~\eqref{eq:sv}.

The former result can also be obtained from the 2-to-4 interaction rate near the kinematical threshold $\sqrt{s} \simeq 4\, m$. Taking into account that the phase-space volume at the threshold is
\begin{equation}
    V = \frac{1}{105 \times 2^{12} \pi^8}\, \sqrt{\frac{m}{2}} \left(\sqrt{s} - 4\,m\right)^{7/2},
\end{equation}
and that the kinematical factor at the threshold is
\begin{equation}
    \frac{(2 \pi)^4}{4\, |\vec p_\text{cm}|\, \sqrt{s}} = \frac{4 \pi^4}{\sqrt{3}\, m^2}\,,
\end{equation}
one finds that the 2-to-4 cross section at the threshold reads
\begin{align}
    \sigma_{2\to 4}(s) &= \frac{4 \pi^4}{\sqrt{3}\, m^2} \times \frac{1}{105 \times 2^{12} \pi^8}\, \sqrt{\frac{m}{2}} \left(\sqrt{s} - 4\,m\right)^{7/2} \times \frac{24^3\, \ls^4}{m^4} \nonumber\\
    &= \frac{9\, \ls^4}{70 \pi^4}\, \sqrt{\frac{m}{6}}\, \frac{\left(\sqrt{s} - 4\, m\right)^{7/2}}{m^6}\,.
\end{align}
This cross section coincides with the numerical value computed with {\tt CalcHEP}.

The thermally averaged 2-to-4 cross section $\svW_{2\to 4}$ is defined as
\begin{align}
    \svW_{2\to 4} &= \frac{1}{8\, m^4\, T\, K_2^2(m/T)} \int_{4 m^2}^\infty \sigma_{2\to 4}(s) \times (s- 4 m^2)\, \sqrt{s}\, K_1(\sqrt{s}/T)\, ds \nonumber\\
    &= \frac{27 \sqrt{3}\, \ls^4}{4 \pi^4}\, \frac{T^3}{m^5}\, e^{-2 m/T},
\end{align}
again near the kinematical threshold. In addition, in the Boltzmann equation for the DM number density, the collision term containing the 2-to-4 and 4-to-2 annihilations has the form
\begin{equation}
    \mathcal{C} = -n^4\, \sv + n^2\, \svW_{2\to 4}
\end{equation}
and therefore at equilibrium
\begin{equation}
    \sv = \frac{\svW_{2\to 4}}{n_\text{eq}^2}\,.
\end{equation}
It follows that the 4-to-2 DM annihilation cross section is
\begin{equation}
    \sv = \frac{54\, \sqrt{3}}{\pi}\, \frac{\ls^4}{m^8}\,,
\end{equation}
which again corresponds to Eqs.~\eqref{eq:sv} and~\eqref{eq:sv2}. We note a 1/16 factor difference with respect to the expression reported in Ref.~\cite{Bernal:2015xba}.

%%%%%%%%%%%%%%%%%%%%%%%%%
\bibliographystyle{JHEP}
\bibliography{biblio}
%%%%%%%%%%%%%%%%%%%%%%%%%
\end{document}